\documentclass[final,authoryear,5p]{elsarticle}

\usepackage{graphicx}

\usepackage{amssymb}

\usepackage[pdftex,%
a4paper=true,%
breaklinks=true,%
colorlinks=true,%
pdfauthor={First Author et al.},%
pdftitle={Template for manuscripts in Advances in Space Research}%
]{hyperref}

\journal{Advances in Space Research}

\begin{document}

\begin{frontmatter}



\title{Cut-off features in interplanetary solar radio type IV emission}


\author{Silja Pohjolainen\corref{cor}}
\address{Tuorla Observatory, Department of Physics and Astronomy, University of Turku, 20014 TURUN YLIOPISTO, Finland}
\cortext[cor]{Corresponding author}
\ead{silpoh@utu.fi}


\author{Nasrin Talebpour Sheshvan}
\address{Department of Physics and Astronomy, University of Turku, 20014 TURUN YLIOPISTO, Finland}
\ead{natash@utu.fi}

\begin{abstract}
Solar radio type IV bursts can sometimes show directivity, so that no
burst is observed when the source region in located far from the solar
disk center. This has recently been verified also from space
observations, at decameter wavelengths, using a 3D-view to the Sun
with STEREO and Wind satellites. It is unclear whether the
directivity is caused by the emission mechanism, by reduced radio wave
formation toward certain directions, or by absorption/blocking of
radio waves along the line of sight. We present here observations of
three type IV burst events that occurred on 23, 25, and 29 July 2004,
and originated from the same active region.  The source location of
the first event was near the solar disk center and in the third event
near the west limb. Our analysis shows that in the last two events the
type IV bursts experienced partial cut-offs in their emission, that
coincided with the appearance of shock-related type II bursts. The
type II bursts were formed at the flanks and leading fronts of
propagating coronal mass ejections (CMEs). These events support the
suggestion of absorption toward directions where the type II shock
regions are located.
\end{abstract}


\end{frontmatter}

\parindent=0.5 cm


\section{Introduction}

Flares and coronal mass ejections (CMEs) are solar transients that can
accelerate particles to high energies and create radio emission within large
wavelength ranges, see reviews in \cite{pick2008} and \cite{nindos2008}.
At low frequencies, below 1 GHz, the radio emission is most often plasma
emission as the trapped or propagating particles cause oscillations in the
surrounding medium and create radio waves that appear at the local plasma
frequency and/or its harmonics. The emission frequency can then be used
as a direct measure of density, and an indirect measure of height, in the
solar atmosphere \citep{white2007}.
Gyrosynchrotron emission is also possible when strong magnetic fields are
present, but to be observable the gyrofrequency must then exceed the local
plasma frequency \citep{dulk85}.

Solar radio type IV bursts were studied intensively in the 1960s and
their main features are described in \cite{kundu65}. The main division was made 
between moving and stationary bursts: the former are moving in solar altitude
and the latter are located low in the corona with no systematic movement
\citep{boischot1957}.
Moving type IVs were estimated to be non-directive and emitting mainly
weakly-polarized synchrotron emission, while stationary bursts were highly
directive toward the center of the solar disk and emitted strongly-polarized
plasma radiation. The cause for stationary type IV bursts could be flare loops,
located high in the corona and filled with non-thermal particles.
Moving type IV bursts would require a propagating CME or an expanding arch,
to create a frequency-drifting structure in the dynamic spectra.  

The fine structures of metric type IV bursts, typically embedded in burst
continua, can be used to trace the magnetic field restructuring and the
corresponding energy release in the low corona \citep{bouratzis2015}. The
observed fine structures include type III-like short duration bursts, zebra
patterns, and stripes with both positive and negative frequency drifts
\citep{melnik08,bouratzis2016}.

\begin{table*}[!ht]
\caption{Flares and CMEs associated with radio type IV bursts.}
\label{table1}              
\begin{tabular}{l l l l c c c c l}
\hline       
Date       & Flare     & GOES    & AR       & Metric type IV  & IP type IV  & IP type II  & CME         & CME        \\
           & start     & class   & location & burst start     & burst $^1$  & burst       & height$^2$   & speed$^3$ \\
yyyy-mm-dd & (UT)      &         &          & (UT)            & (UT)        & (UT)        & (R$_{\odot}$) & (km s$^{-1}$)\\
\hline      
2004-07-23 & 17:07     & M2.2    & N03W04   & 19:10          &19:30--21:00  & 18:30--19:30 & 8.0       & CME1:  600\\
           & 18:02     & C4.1    & N05W05   &                &              &              & 4.5       & CME2:  550\\
2004-07-25 & 13:37     & M2.2    & N04W30   & 14:20          &15:10--20:00  & 15:30--19:00 & 4.7       & CME1:  450\\
           & 14:19     & M1.1    & N08W33   &                &              & 17:40--      & 4.3       & CME2: 1300\\
2004-07-29 & 11:42     & C2.1    & W-limb   & 11:20          &12:40--14:30  & 13:30--      & 4.5       & 1000\\
\hline                  
\end{tabular}\\
$^1$Start time defined as appearance at 14 MHz (Wind/WAVES upper frequency limit) .\\ 
$^2$CME leading front height at the time of IP type IV appearance (from fits to the height data if not observed).\\
$^3$CME speed at the time of IP type IV appearance. 
\end{table*}

Before space instruments the observable frequency range for type IV bursts
was limited to meter-decameter waves by the Earth atmosphere, but since
the early 1990s radio dynamic spectra have been available from
decameter-hectometer (DH) to km waves. Observations at DH wavelengths probe
the plasma conditions and transients in the interplanetary (IP) space, and
currently radio spectra are available from the spectrographs on board
Wind \citep{bougeret95} and STEREO \citep{bougeret2008} satellites.

IP type IV bursts observed by the Wind/WAVES in 1998-2012 have
been listed by \cite{hillaris2016}. They determined the bursts to be moving
type IVs on the basis that their emission drifted to DH wavelengths. However,
as there is no radio imaging available at DH waves at present, their actual
movement in altitude is difficult to verify. 

The directivity in IP type IV bursts was first reported by \cite{gopal2016}.
By comparing the radio spectra observed from three different viewing angles
around the Sun (STEREO A and B, and Wind), they observed that the IP 
type IV bursts are visible only if their source location is within $\sim$60
degrees from the Sun center. Their conclusion was that the type IV emission
is directed along a narrow cone above the flare site, the emission
mechanism is most probably plasma emission, and the type IV bursts are 
stationary, i.e., not associated with moving CME structures. However, in a
later case study \cite{melnik18} were able to show that the IP type IV burst
that was not observed by Wind near Earth (but visible in both STEREO A and B
dynamic spectra), could still be observed from Earth at decameter waves,
by URAN-2. 

Recently, \cite{nasrin2018}, who used similar data from the three spacecraft,  
suggested that the observed directivity of IP type IV bursts may not be due
to the emission directivity, but could be caused by absorption or suppression
of emission toward certain viewing angles. The absorber, i.e., dense plasma
along the line-of-sight, would block the emission at longer DH wavelengths.
Such high-density regions could be formed by shock-streamer interaction,
especially near CME flanks. 

This article presents the analysis of three separate eruption events that
occurred in July 2004 (before STEREO was launched and therefore only Earth
view available). These events all included flares, CMEs, type II bursts, and
metric and DH type IV bursts. They originated from the same active region,
which rotated from a location near the center of the disk (first event)
to the west limb (third event), and thus made it possible to observe the
type IV bursts and the associated features near the solar disk center and
near the limb. 

In the analysis we used radio spectral data from Wind satellite, Radio Solar
Telescope Network (RSTN) stations, Green Bank Solar Radio Burst Spectrometer
(GBSRBS), and Nancay Decameter Array (NDA). Nancay Radioheliograph (NRH)
images were used to locate the burst sources, but these observations were
available only for the two European daytime events. SOHO/LASCO coronagraph
data and SOHO/EIT images in EUV were taken from the CDAW CME Catalog
at \url{https://cdaw.gsfc.nasa.gov/}.

\section{Data analysis: solar events in July 2004}

In July 2004 the NOAA solar Active Region (AR) 10652 produced three eruption events
that were associated with flares, CMEs, and radio type II and type IV
bursts. The events occurred on 23 July (eruption source location at N05W05),
25 July (source location at N04W30), and 29 July (source at west limb).
The flares had moderate intensities, with GOES X-ray classes ranging from C2.1 to M2.2.
The CMEs were observed in projection, which may affect their height estimates 
especially when the CME is propagating toward the observer. Fits to the height 
estimates gave CME speeds that ranged from 450 km~s$^{-1}$ to 1300~km~s$^{-1}$, see
Table~\ref{table1}.   

In all three events both a metric type IV burst (ground-based observations at
frequencies higher than 20 MHz) and an IP type IV burst (space observations at
frequencies lower than 14 MHz) were observed, and they look to be related, i.e.,
the IP emission was a continuation of the metric emission. 

The heights of plasma emission sources can be estimated using solar atmospheric
density models, as the frequency of plasma emission depends only on the local
plasma density, see, e.g., \cite{pohjolainen2007}. The 'hybrid' atmospheric
density model of \citet{vrsnak04} was selected for our analysis because it was
developed to merge the high-density low corona models to the low-density IP
models without breaks or discrepancies, which fits the purpose of our analysis. 

\begin{figure*}[!ht]
   \centering
   \includegraphics[width=0.4\textwidth]{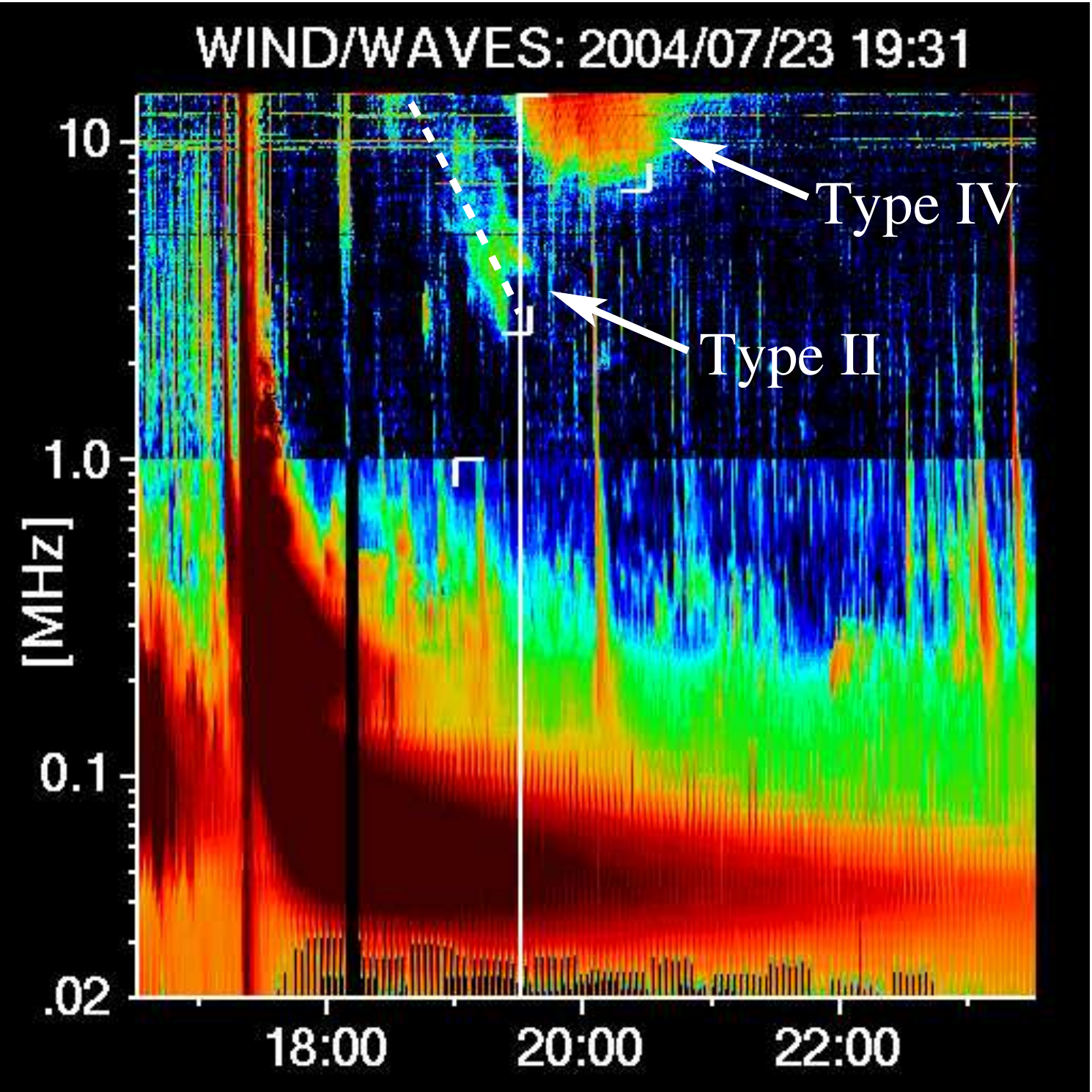}
   \includegraphics[width=0.55\textwidth]{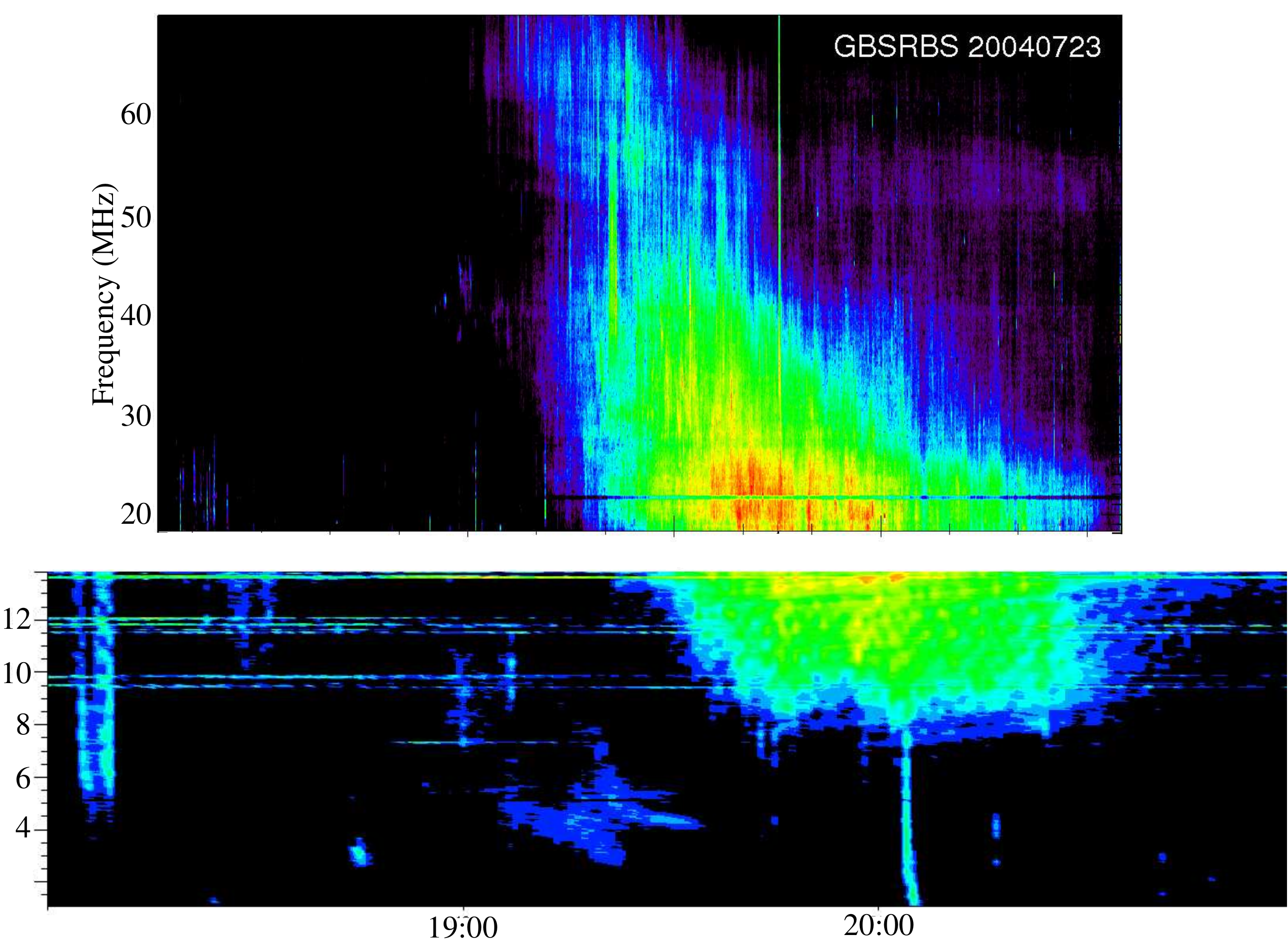}
   \includegraphics[width=0.4\textwidth]{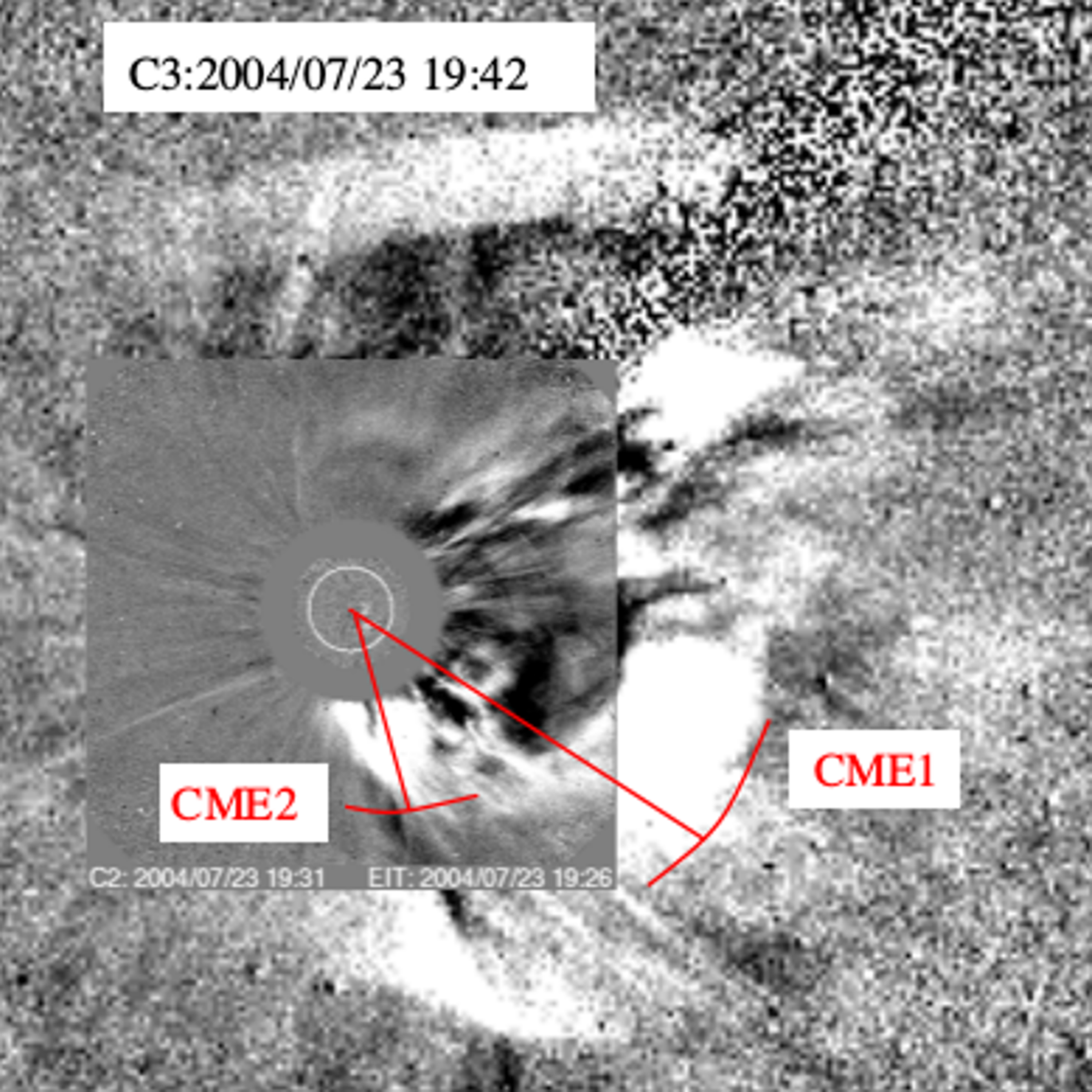}
   \includegraphics[width=0.55\textwidth]{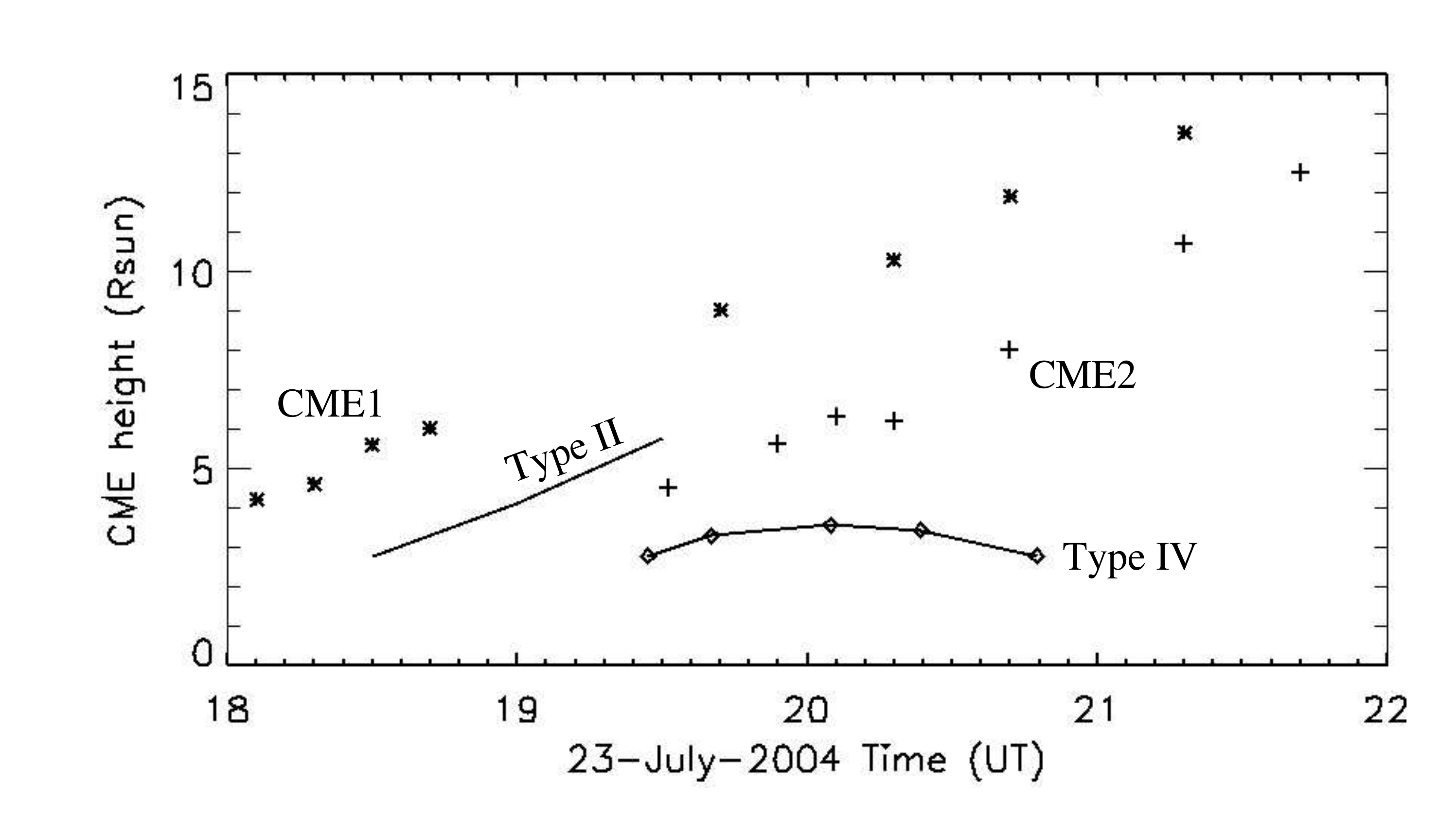}
   \caption{Solar event on 23 July 2004, with source location at N05W05.
     The type II burst is indicated by a dashed line in the Wind/WAVES dynamic
     spectrum (top left).
     The metric type IV burst (GBSRBS) continues to DH wavelengths (Wind/WAVES),
     their dynamic spectra are both plotted in linear scale (top right).
     The event included two separate CMEs, shown in the combined LASCO C2/C3
     difference image near the time of the type II burst disappearance (bottom left).
     Three streamers were observed on the western side of the Sun, and one of them was
     located near the southern flank of CME1. 
     The height-time plot (bottom right) shows the estimated type II burst
     heights (solid line), the estimated type IV burst heights (connected triangles),
     and the projected CME heights (stars and crosses for the two CMEs).}
\label{july23}%
\end{figure*}

\begin{figure*}[!ht]
   \centering
   \includegraphics[width=0.36\textwidth]{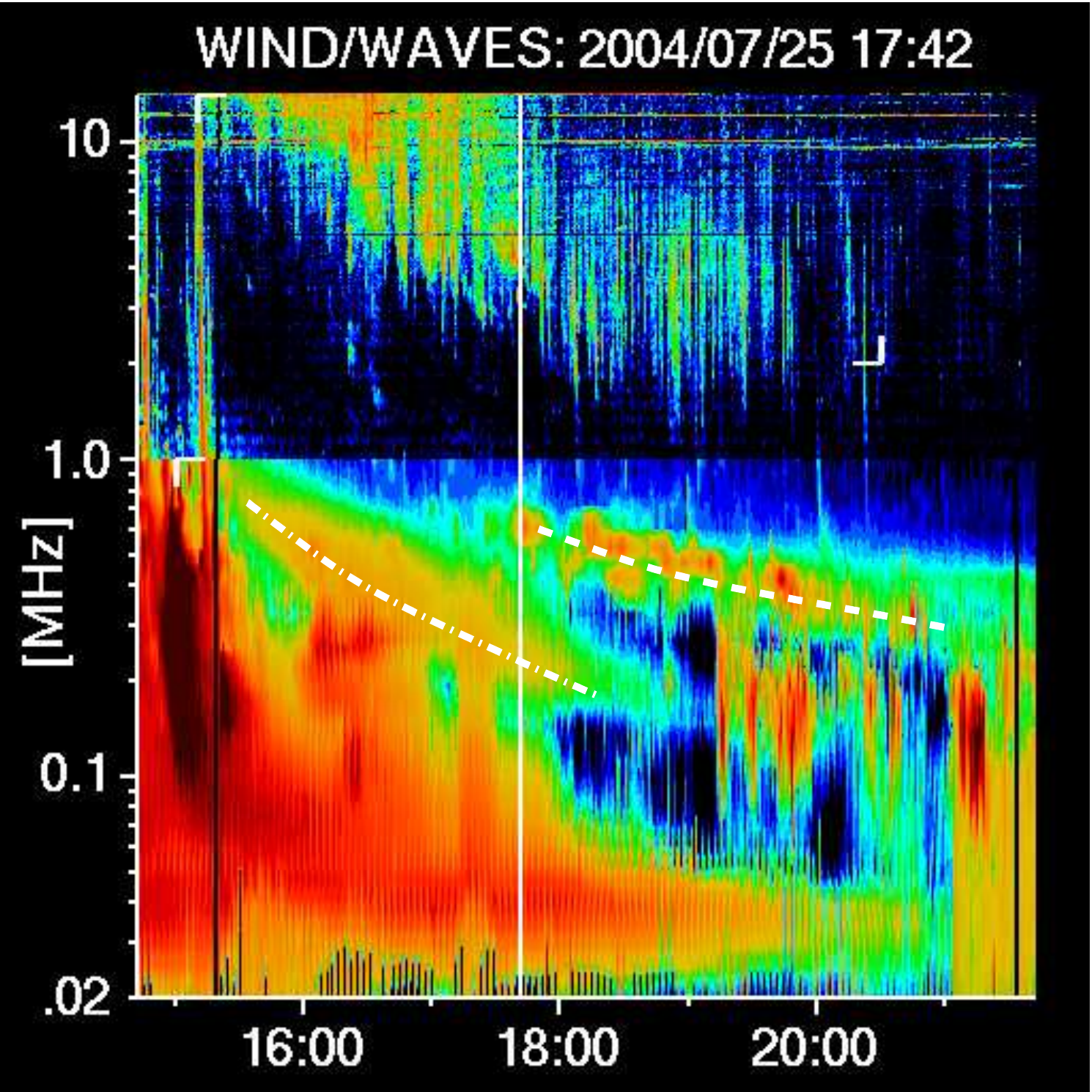}
   \includegraphics[width=0.55\textwidth]{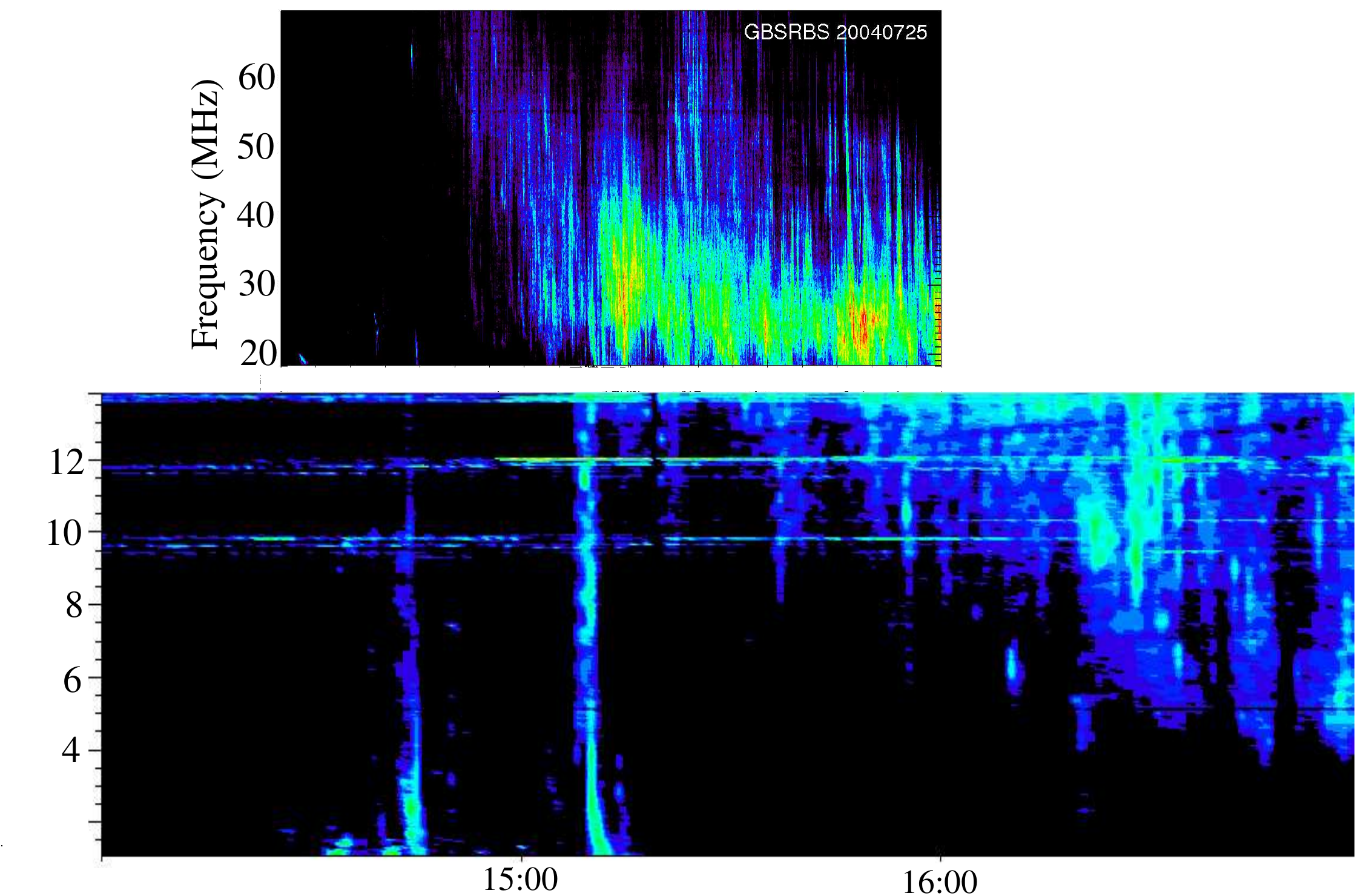}
   \includegraphics[width=0.36\textwidth]{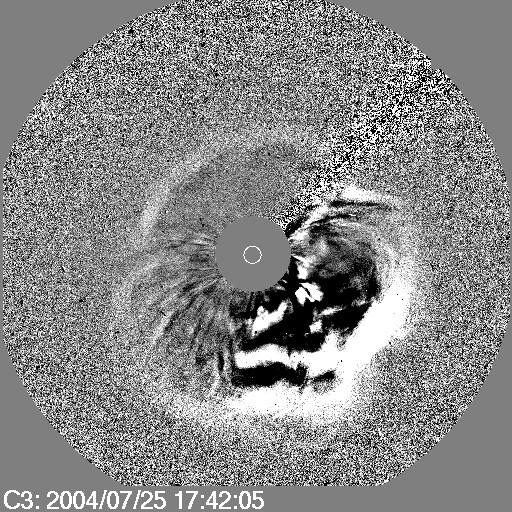}
   \includegraphics[width=0.55\textwidth]{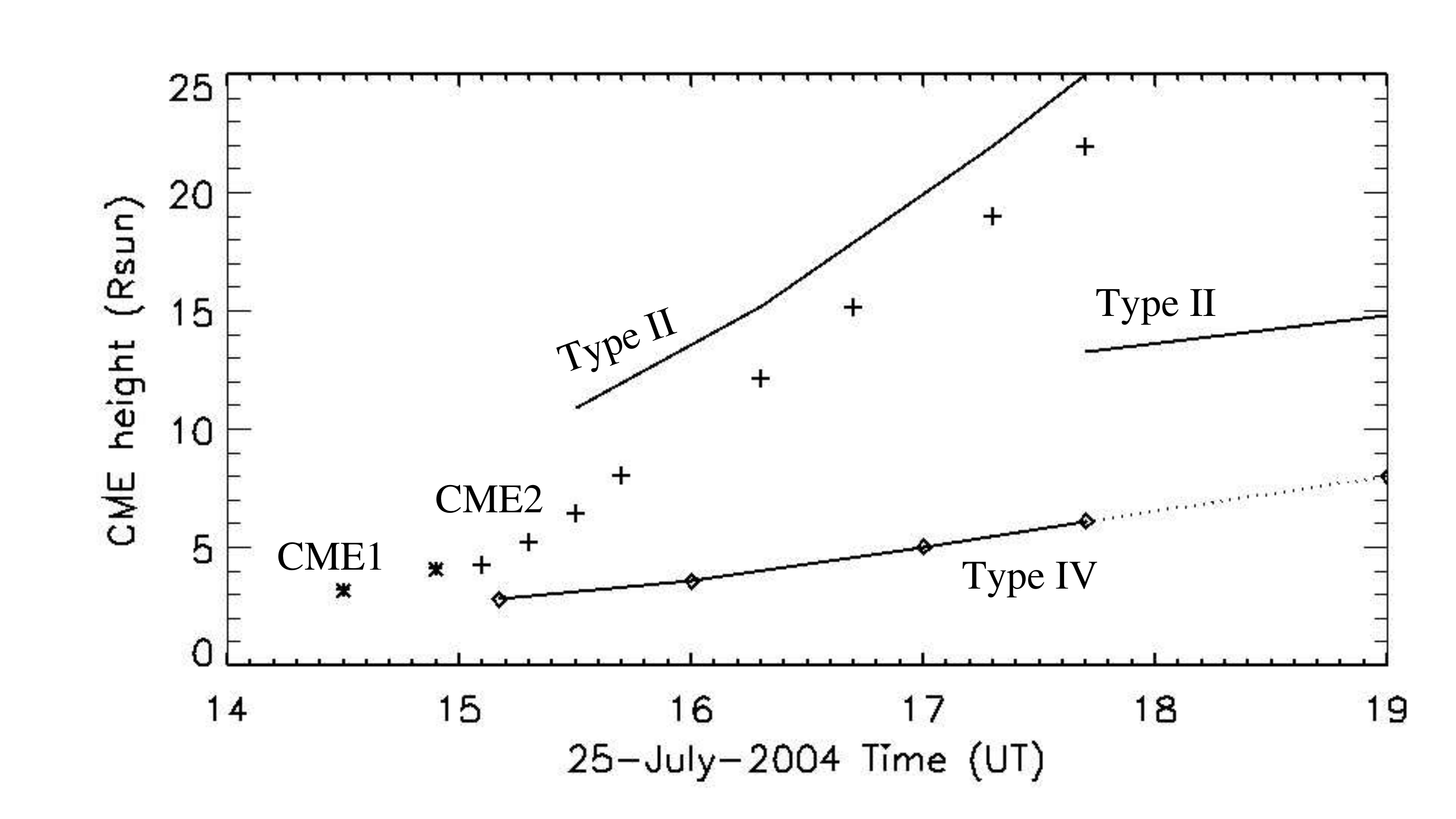}
   \caption{Solar event on 25 July 2004, with source location at N04W30.
     The first-appearing wide-band type II burst is indicated by a dash-dotted line
     and the later narrow-band type II burst by a dashed line in the Wind/WAVES
     dynamic spectrum. The vertical white line indicates the time when the type IV
     burst intensity started to decrease (top left). 
     The metric type IV burst (GBSRBS) continues to DH wavelengths (Wind/WAVES),
     their dynamic spectra are plotted in linear scale (top right).
     The merged CME is shown in the LASCO C3 difference image near the time of the
     second type II burst appearance and the partial disappearance of the type IV burst
     (bottom left). Streamers were observed both in the north-west and south-west directions.
     The height-time plot (bottom right) shows the estimated type II burst heights
     (solid lines), the estimated type IV burst heights (connected triangles), and
     the projected CME heights (stars and crosses, the CMEs merge as one near 15 UT).}   
\label{july25}%
\end{figure*}

\subsection{Event on 23 July 2004}

The event originated near the disk center, at $\sim$ N05W05, and was associated
with continuous flaring from AR 10652 (see Table \ref{table1}).
Two CMEs were launched from this region within the time period of 17--19 UT,
the first one (CME1) was a partial halo-type and the second one (CME2) had
more narrow width. They were preceded by a halo CME that was last observed at
17:42 UT at a height of 9.5~R$_{\odot}$. It is unclear if the earlier halo CME
merged or interacted with CME1, which was first observed at 17:54 UT at
height 3.2~R$_{\odot}$.

A metric type IV burst was observed to start at 19:10 UT, observed by GBSRBS and
RSTN Palehua stations, and the emission drifted to decameter waves where it was
observed by Wind/WAVES. Only one IP type II burst was observed at 18:30--19:30 UT,
and its emission ended soon after the appearance of the type IV burst at WAVES
frequencies. A faint metric type II burst was observed earlier, at 16:00--16:15 UT,
but it was most probably related with the acceleration of the halo-type CME observed
before CME1 and CME2. 

The height-time plot in Figure~\ref{july23} shows that the estimated height of the
IP type II burst source (the type II lane is indicated with a dashed line in the top
left panel) was 2--3~R$_{\odot}$ lower than the (projected) leading front of CME1.
The CME2 heights cannot be determined accurately before 19:31~UT, but a
filament eruption in the active region is visible in the SOHO/EIT images at 18:48~UT.
As type II bursts often appear at the time of CME acceleration, it is possible that
if the CME2 was launched near 18:30~UT and it was accelerating, it could have
been the source of the type II burst. Furthermore, as CME heights are difficult to
determine for on-the-disk events, as CMEs are seen as projections, the true CME2 heights
could have been higher than what was observed. On the other hand, CME1 propagated
toward the south-west and a streamer was located at its southern flank, suggesting
that a CME flank shock would have been possible, too.

The type II burst disappeared from the spectrum near 19:30 UT, having reached
3 MHz frequency, which corresponds to a source height of about 5.7~R$_{\odot}$.
The type~II burst source was then $\sim$1~R$_{\odot}$ higher than the leading
front of CME2 and $\sim$2~R$_{\odot}$ lower than that of CME1. Near this time both
CMEs had relatively low speeds, 550--600~km~s$^{-1}$, and the abrupt disappearance of
the type II burst lane may simply have been caused by the shock speed falling below
the local Alfven speed. Or, as in  \cite{kong2015}, the type II emission ended once
the CME/shock front passed the white-light streamer tip.  

The type IV burst appeared in the Wind/WAVES spectrum at 14 MHz at about 19:30 UT.
Figure~\ref{july23} shows the estimated type IV burst heights that were calculated
from the lowest emission frequencies of the type IV emission (leading emission
structure), assuming that the frequency drift corresponds to the decreasing plasma
density in the surrounding medium.

\subsection{Event on 25 July 2004}

The long-duration M1.1 flare at N08W33 started after a series of impulsive compact
flares that occurred over a period of two hours. The soft X-ray images analyzed
by \cite{vrsnak2010} revealed that the eruption formed a large cusp-shaped loop
system in between AR 10652 and AR 10653, which was located south of AR 10652. 
The launched CME was a full halo-type and had high speed and almost constant velocity,
1300 km s$^{-1}$, after 15 UT. A slower CME was observed to precede this CME, possibly
merging with it near 15 UT (Table \ref{table1} and height-time plot in Figure \ref{july25}). 

A metric type IV burst was observed by GBSRBS, RSTN San Vito station, and NDA, and the
emission drifted to decameter waves where it was observed by Wind/WAVES. NDA observations
show that the burst was only very weakly left-hand polarized. (The NDA spectra are
available at the Radio Monitoring website, \url{http://secchirh.obspm.fr/}).
The type IV burst drifted to the WAVES upper frequency limit, 14 MHz, at 15:10 UT.

At 15:30 UT a wide-band type II emission lane appeared at 1 MHz (indicated with a
dash-dotted line in Figure~\ref{july25}), and a second, more narrow-band and
fragmented type II lane at 17:40 UT at a higher frequency (dashed line
in Figure~\ref{july25}). The IP type IV burst showed a clear intensity decrease
after 17:40 UT, and the wide-band type II emission ended near 19:00 UT. 

The calculated heights for the first, wide-band type II burst suggest that it was
due to a CME bow shock, as the radio emission sources were located just above
the projected heights of the CME leading front. This burst has been analyzed earlier
by \cite{pohjolainen2013}, as part of a large sample of wide-band type II bursts,
where the majority were concluded to be CME bow shocks.

The second, more narrow-band type II burst appeared at 17:40 UT at higher
frequencies, and had therefore much lower heights than the wide-band type II
burst. The calculated speed of the second shock is also lower (see height-time
plot in Figure~\ref{july25}), as it would be if the shock was located at the
flanks of the CME instead of the faster-moving CME nose. The partial blocking
of the type IV burst emission, which also started at 17:40 UT, suggests a
connection to the second shock.

\begin{figure*}[!ht]
   \centering
   \includegraphics[width=0.39\textwidth]{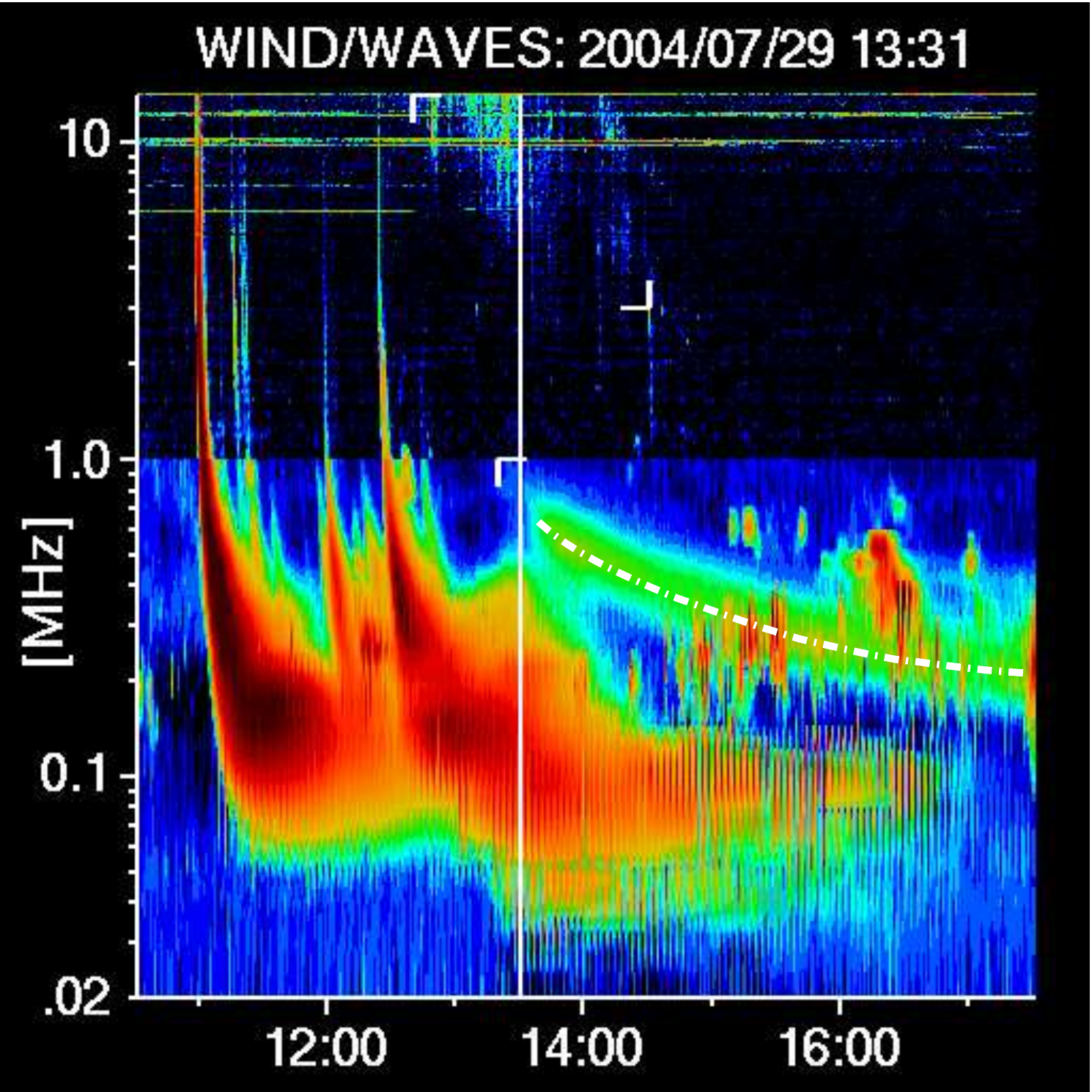}
    \includegraphics[width=0.51\textwidth]{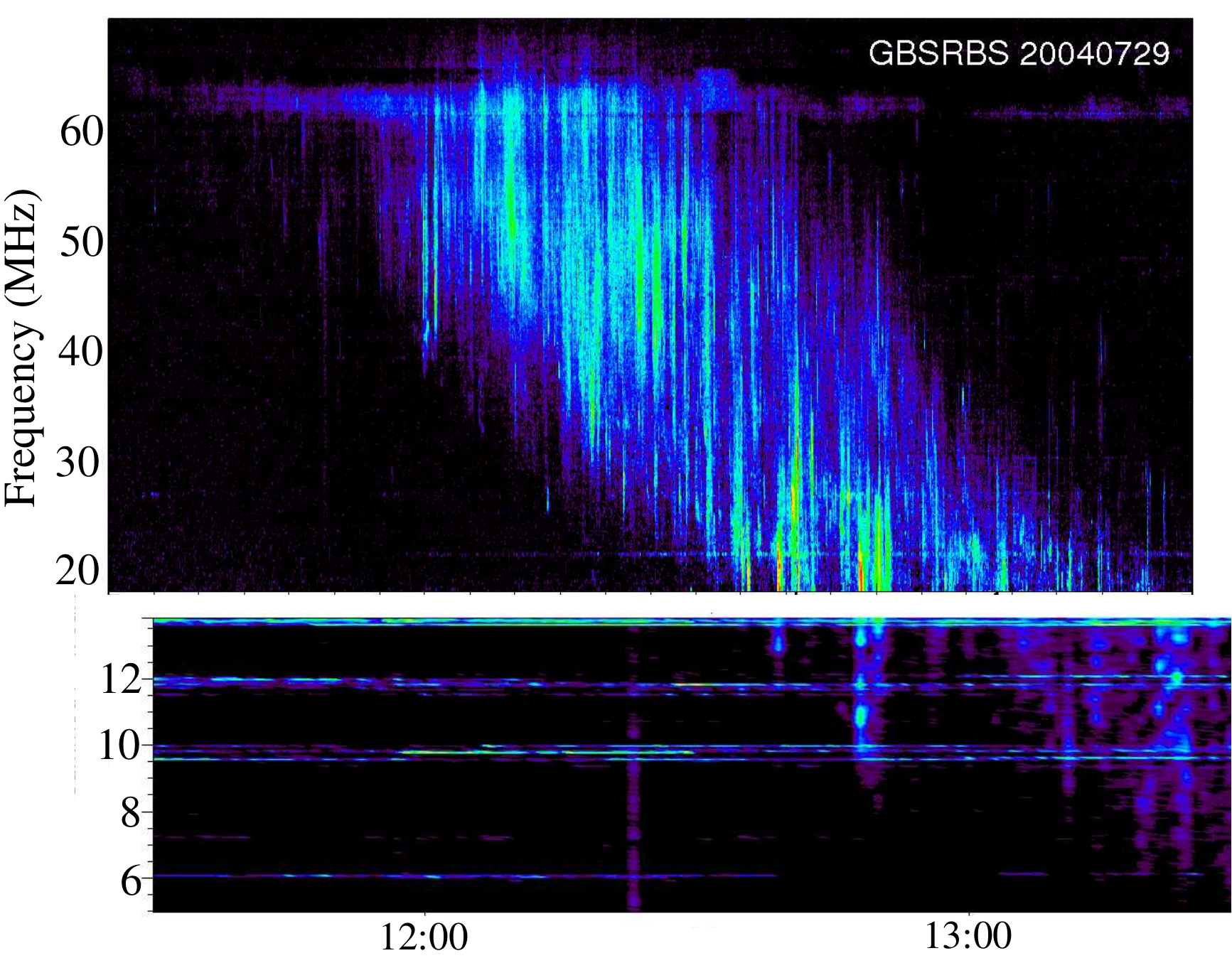}
   \includegraphics[width=0.39\textwidth]{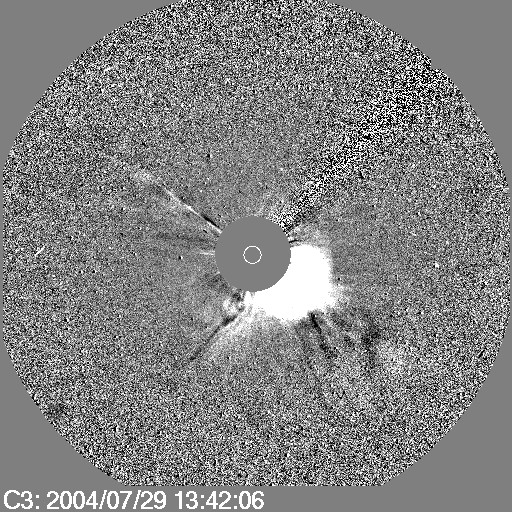}
   \includegraphics[width=0.51\textwidth]{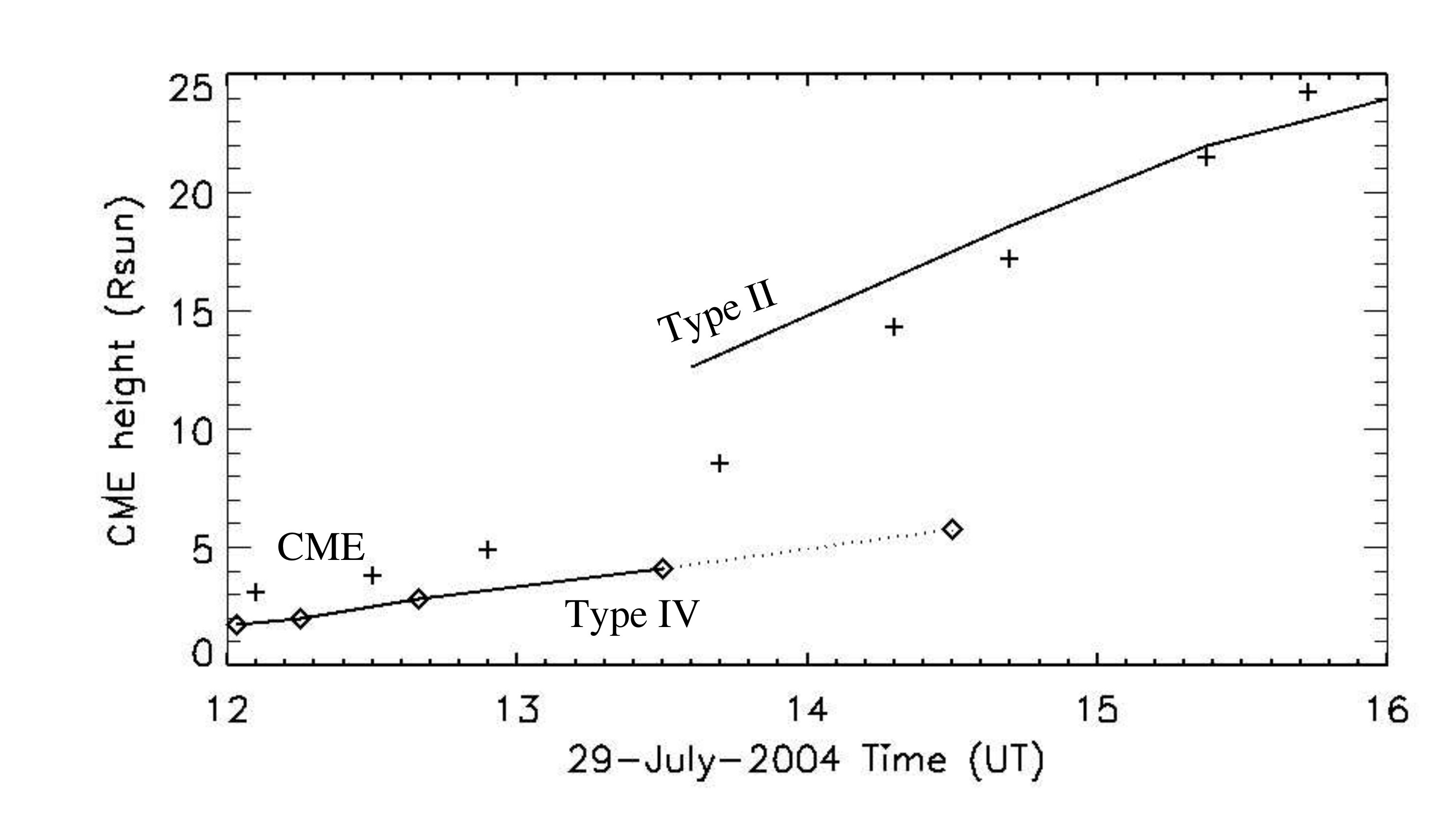}
   \caption{Solar event on 29 July 2004, with source location near the west limb.
     The wide-band type II burst is indicated by a dash-dotted line in the
     Wind/WAVES dynamic spectrum. The vertical white line indicates the time
     when the type IV burst intensity started to decrease (top left).
     The metric type IV burst (GBSRBS) continues to DH wavelengths (Wind/WAVES),
     their spectra are plotted in linear scale (top right).
     The CME is shown in the LASCO C3 difference image near the time of the type II
     burst appearance and the partial disappearance of the type IV burst (bottom left).
     Streamers were observed both in the north-west and south-west directions.
     The height-time plot (bottom right) shows the estimated type II burst heights
     (solid line), the estimated type IV burst heights (connected triangles), and
     the projected CME heights (crosses).}
   \label{july29}%
\end{figure*}

\subsection{Event on 29 July 2004}

On 29 July 2004 AR 10652 had rotated to the west limb. A long-duration X-ray
flare was observed, associated with a high-speed (1000 km s$^{-1}$) halo-type CME.
 \cite{liu2010} report that the halo CME consisted of a transequatorial
loop system that connected AR 10652 to AR 10653, and the loop system was mainly
oriented in the north-south direction.
The GOES C2.1 flare is listed to start at 11:42 UT, but there is strong
indication that the flare had started earlier on the far side of the Sun.
  
A type IV burst was formed at meter waves at 11:20~UT, and the burst envelope
consisted of narrow-band type III-like bursts and fluctuations. 
The burst envelope drifted to DH wavelengths where Wind/WAVES observed it
(Figure~\ref{july29}) until 14:30 UT, down to $\sim$3 MHz.
\cite{liu2010} defined this spectral feature as a drifting
pulsating structure (DPS) that would be formed within the current sheet below
the CME, see their Figure 2. We note that as DPS's are typically observed at
decimeter waves, at much lower heights, this interpretation may be questioned.
NDA observations show that the type IV burst was only very weakly left-hand polarized.

Only one wide-band DH type II lane was observed (Figure~\ref{july29}, top left
dynamic spectrum, indicated by a dash-dotted line) and it appeared more than 
two hours after the start of the metric type IV emission (see Table \ref{table1}).

The height-time plot in Figure \ref{july29} shows that the type~II burst was most
likely due to a bow shock at the leading front of the CME, initiated by a change
in the CME velocity around 13:30 UT. (Note that in the LASCO CDAW catalog the
LASCO C3 image at 13:42 UT has not been included in the fits to the CME height
data.) Details of the wide-band type II burst and comparison to the CME height data
can also be found in \cite{pohjolainen2013}.  

Near the time of the wide-band type II burst appearance, the type IV burst started
to fade away in the dynamic spectrum, after which it disappeared gradually. 

\subsection{Type IV burst features}

\begin{figure*}[!ht]
  \begin{center}
  \includegraphics[width=0.331\textwidth]{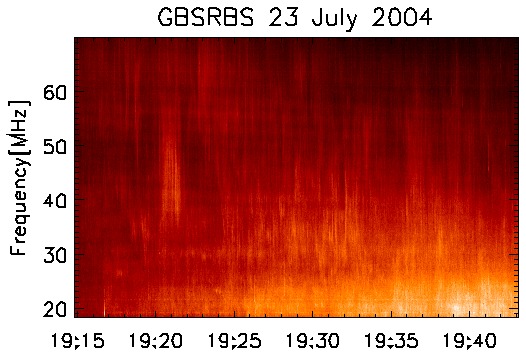}
  \includegraphics[width=0.318\textwidth]{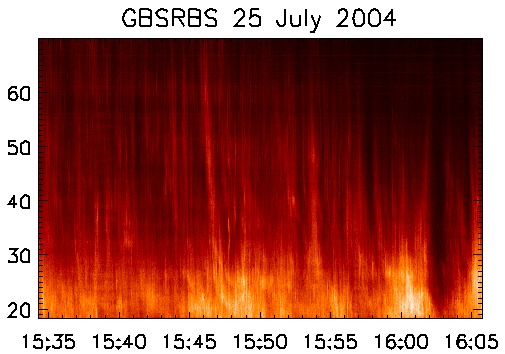}
  \includegraphics[width=0.312\textwidth]{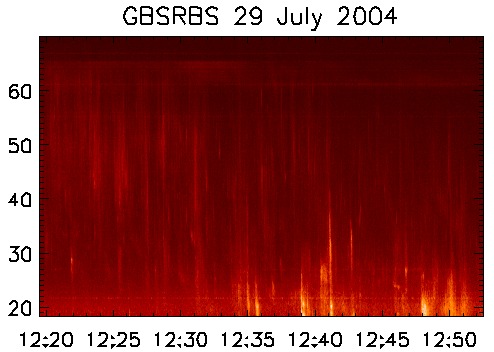}
  \end{center}
  \caption{GBSRBS dynamic spectra at 70--20 MHz from the three events 
    show fine structures within the metric type IV continuum emissions.}
  \label{gbsrbs}%
\end{figure*}

The three metric type IV bursts showed similar fine structures inside a diffuse
frequency-drifting emission envelope. The fine structures included type III-like
bursts that appeared in groups (Figure~\ref{gbsrbs}). A simple Lomb-Scargle periodicity
test revealed that some of the burst groups were repeated every $\sim$ 100 s.
As type III bursts are generated by particle beams traveling along magnetic field lines,
this indicates continuous particle acceleration during the type IV bursts.
Hard X-ray observations were available only for the 23 July event (RHESSI satellite data)
and they showed X-ray burst activity during 19:15--19:50 UT, especially in the 6--50 keV
energy range, but no one-to-one correspondence to the type III bursts were found.

Both positive and negative frequency drifts were observed in the radio spectral
data at metric wavelengths. The typical bandwidths of these features
were some tens of MHz, indicating that the particle beam paths were confined to
relatively short distances. The Wind/WAVES spectral data did not have enough
time resolution to determine if these fine structures existed also at longer wavelengths.
Positive and negative frequency drifts suggest that particles were trapped inside
large loops (flare or CME), where they moved along the magnetic field lines toward
both lower and higher density regions.

\begin{figure}[!h]
  \begin{center}
  \includegraphics[width=0.45\textwidth]{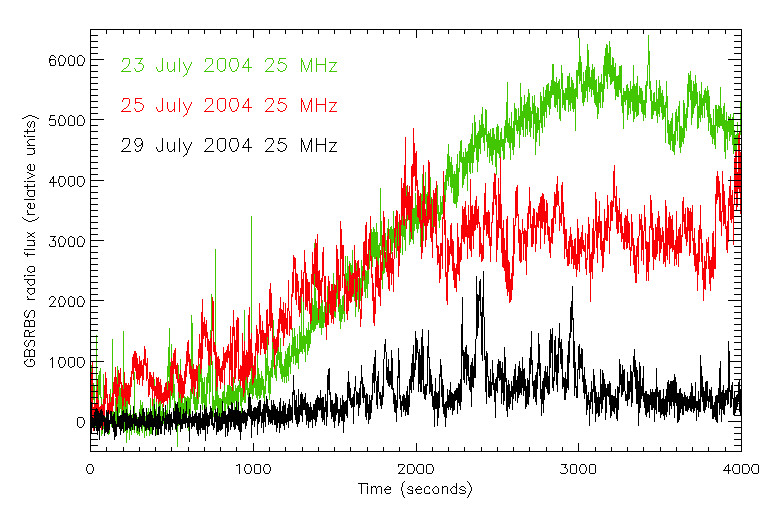}
  \end{center}
  \caption{Background subtracted GBSRBS radio flux at 25 MHz in
    the three type IV burst events.}
  \label{cont}%
\end{figure}

We note that the diffuse emission inside the envelope was strong\-est in intensity
in the 23 July type IV burst, moderate in the 25 July type IV burst, and relatively
weak in the 29 July type IV burst, see Figure~\ref{cont}.

\subsection{Radio imaging}

\begin{figure}[!h]
  \begin{center} 
\includegraphics[width=0.45\textwidth]{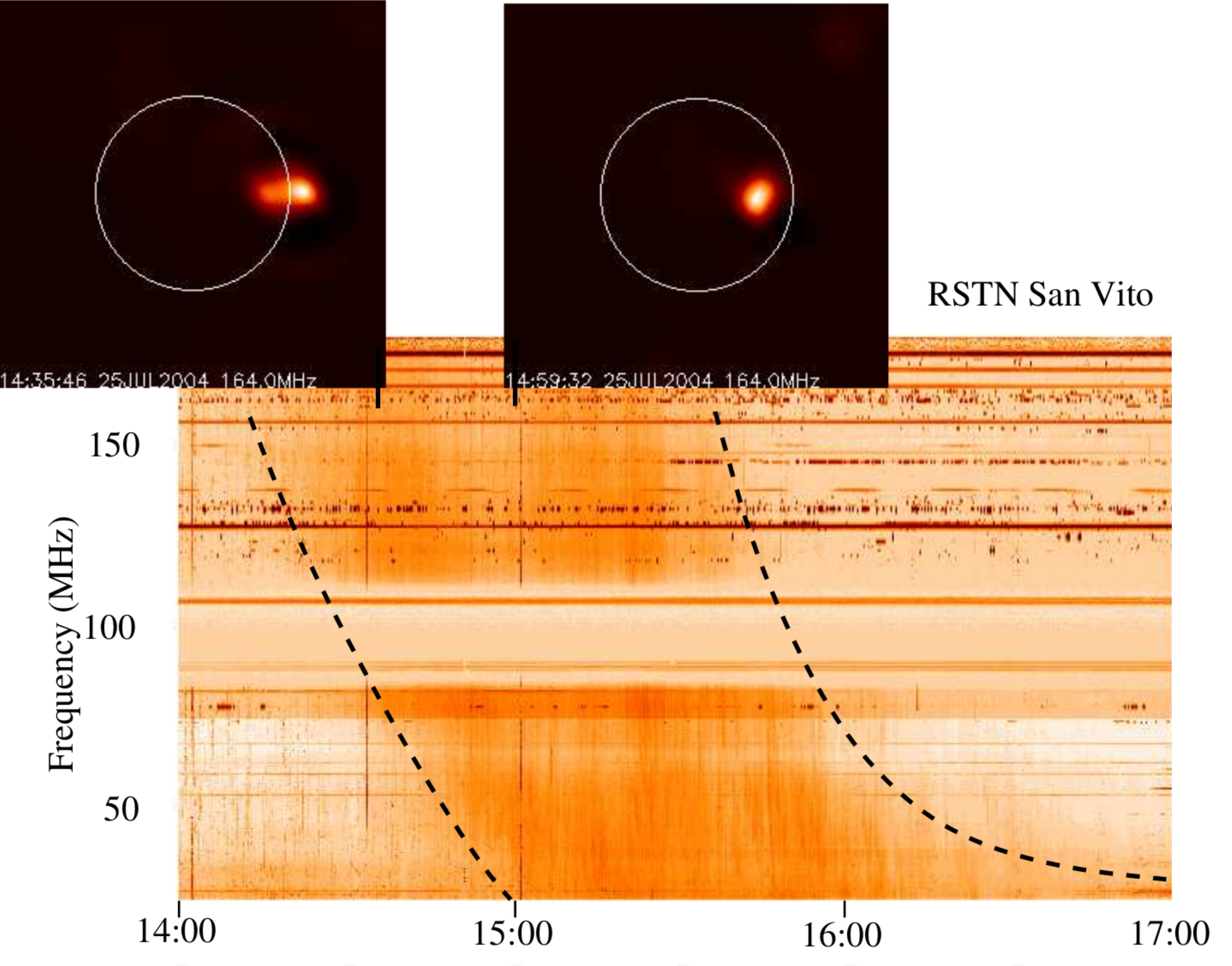}
  \end{center}
\caption{Metric type IV emission recorded by RSTN San Vito on 25 July 2004, in the
  frequency range of 180--25 MHz. Nancay Radioheliograph imaged the radio
  emission sources at 164 MHz, and those are shown at 14:35 UT and 14:59 UT.
  The type IV burst source  was first located near the west limb and then
  moved toward the south-east, in between these images.}
  \label{rstn-25july}%
\end{figure}

The events on 25 and 29 July 2004 were imaged by the Nancay Radioheliograph (NRH)
at 164 MHz. The NRH images on 25 July (Figure~\ref{rstn-25july}), taken during the
time period of the metric type IV emission at 164 MHz, show that the emission
source was first located near the western limb but then moved to the south-east.
As it was discovered earlier \citep{vrsnak2010},
a large cusp-shaped post-eruption loop system was formed south of AR 10652,
connecting it to AR 10653. The radio source movement may therefore have
been due to these post-eruption structures located high in the corona, observed
projected on the solar disk.

\begin{figure}[!h]
  \begin{center}
\includegraphics[width=0.3\textwidth]{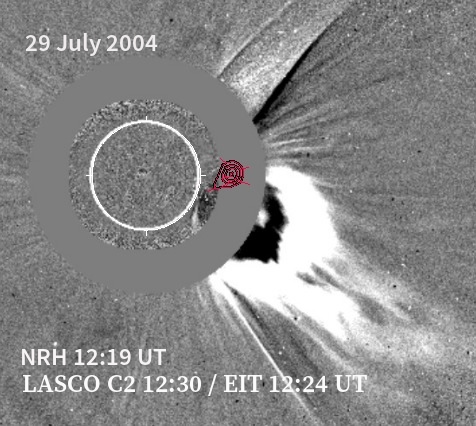}
  \end{center}
  \caption{Nancay Radioheliograph (NRH) observation of the radio emission source
    at 164 MHz at 12:19 UT on 29 July 2004 (red contours) plotted over the SOHO/LASCO
    C2/EIT difference image near that time. The metric type IV burst source at 164 MHz
    is located over the west limb, seemingly near the north-western flank of the CME.}
  \label{rstn-29july}%
\end{figure}

In the 29 July event the metric type IV burst source at 164 MHz was located over
the west limb (Figure \ref{rstn-29july}), seemingly near the north-western flank
of the CME. We estimated that the heliocentric distance of the emission source
was $\sim$1.6 R$_\odot$ at 164 MHz. Plasma emission at this frequency, in typical
atmospheric conditions and densities, is formed near 1.2--1.3 R$_\odot$, so the
emission may not have been plasma emission.

Imaging observations of a moving type IV burst, using the multifrequency
radioheliograph at Clark Lake Radio Observatory, at 125--20 MHz, suggested
that a plasmoid containing energetic electrons can result from the disruption of a
coronal streamer \citep{kundu87}. Furthermore, \cite{gopal90} concluded that as
moving type IV bursts primarily depend on non-thermal particles trapped in moving
magnetic structures (plasmoids and loops that accompany the CMEs), non-thermal
particles can be generated independent of the CME speeds.
A particular kind of type IV bursts, expanding CME radio loops, have
been imaged by, e.g., \cite{bastian2001} and \cite{maia2007}.
The fast-moving radio loops have been observed to move from a few tenths
to more than 1 R$_\odot$ above the solar limb. These features have been attributed
to incoherent synchrotron emission from electrons with energies around 1 MeV.

\section{Summary and discussion} 

We have presented here observations of three different radio type IV burst events
that occurred on 23, 25, and 29 July 2004, and originated from the same active region.
The source location of the first event was near the disk center and in the
third event near the west limb. Our analysis shows that in the last two events
the type IV bursts experienced partial cut-offs in their emission, which 
coincided with the appearance of shock-related type II bursts.

\begin{figure*}[!ht]
  \begin{center}
   \includegraphics[width=0.8\textwidth]{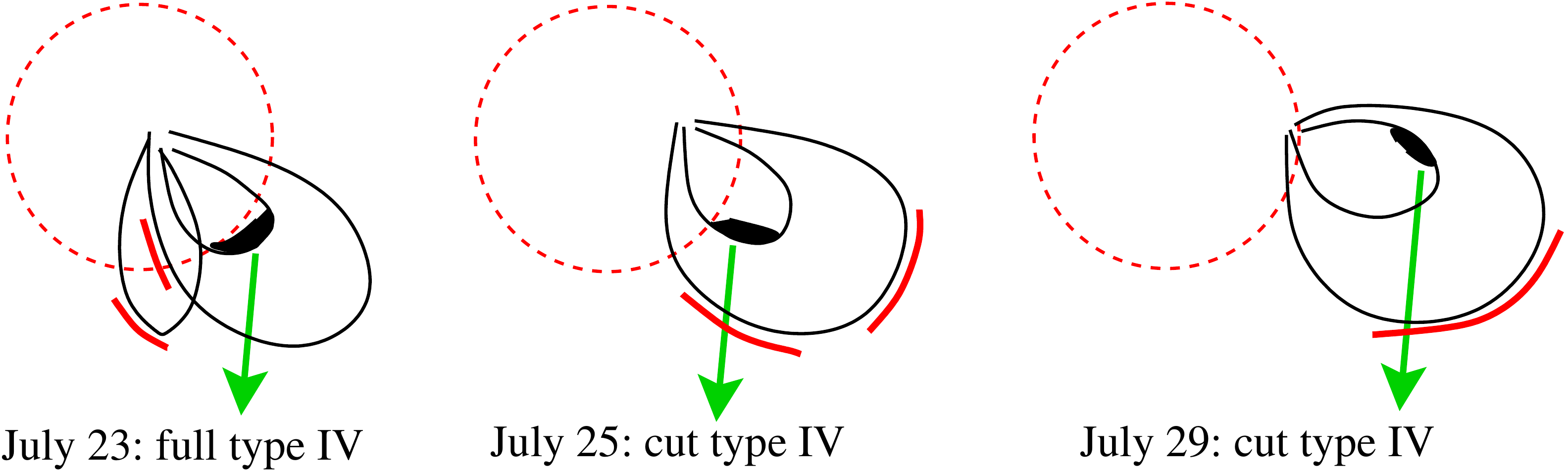}
  \end{center}
   \caption{Schematic cartoons of the three events which show the suggested
     locations of the type II bursts (indicated by red lines) and the type IV
     emitting regions (filled black areas) inside the CME loops (not to scale,
     the radio emission sources could be several R$_{\odot}$ from the Sun center).
     Green arrows show the direction toward the observer and where the type IV
     burst emission may cross the type II shock region.}
 \label{models}%
\end{figure*}

In the 23 July 2004 event the short-duration type II burst was most probably
caused by a shock at the flank of the first CME, or it was due to a bow shock
near the leading front of the second CME.
In a disk-centered event, a CME flank shock would not have been located
in between the type IV burst and the observer, but for a bow shock at the CME front
this would have been possible as the true CME heights could have been larger
than the observed projected heights. In the 23 July event the IP type II burst
ended near the time of the IP type IV burst appearance and the type IV burst 
emission was observed as a typical, uncut envelope structure.

The 25 July 2004 event consisted of two separate type II bursts, of which
the first, wide-band one was most likely formed by a CME bow shock since the
estimated shock heights were well above the projected CME heights.
The second, more narrow-band and slower shock that appeared in the radio
spectrum two hours later was most probably due to a shock at the CME flanks,
at relatively large heights.  
In principle, as two CMEs were observed to propagate in the same direction in
the early stages of the event, the later shock could have been formed also inside
the observed first CME. But, typically type II bursts that are associated
with interacting and merging CMEs show fragmented and enhanced emission, as the
densities in the wake of the first CME vary considerably. As this was not
observed, the flank shock scenario looks to be more probable in this case.
The partial disappearance of the IP type IV burst emission coincided with the
appearance of this second, flank-shock related type II burst.

In the 29 July 2004 event the type IV burst also disappeared partially,
near the time of the appearance of the wide-band type II burst, which was
also determined to be due to a CME bow shock. In both events, on 25 and 29 July,
the type II shock was most probably located in between the observer and
the type IV burst source. Hence, the shock region could have absorbed or
suppressed the type IV emission along this line of sight.

Based on these findings, we present a schematic cartoon in Figure \ref{models},
which shows the configurations in the three events. We propose that in the
two events, on 25 and 29 July 2004, the type IV emission was partially blocked
toward certain viewing angles by shock-related dense plasma. 

The type IV bursts on 25 and 29 July were only weakly left-hand polarized (no
polarization data was available for the 23 July event). This suggests that they
were the moving type, non-directive, and emitted synchrotron radiation. The
imaging of the type IV source region on 29 July also supports synchrotron as
the emission mechanism. The source was located higher in the corona that would
have been possible for a plasma emission source, within typical atmospheric
densities.

Recently \cite{vasanth19} have presented analysis of a similar event,
where the moving type IV emission source was also located very high in the corona,
showed only weak polarization, but had too high brightness temperature to support
the synchrotron radiation mechanism. Therefore \cite{vasanth19} suggested
coherent plasma emission, excited by energetic electrons trapped within the
top front of the bright core of a CME. They state, however, that further studies
are needed to verify the emission mechanism.  

The fine-structures inside the type IV bursts were similar in all our events,
and they showed type III-like, narrow-band fast-drift bursts within the
frequency-drifting diffuse emission envelope. Unfortunately the time resolution
in the radio spectral data was not high enough to verify if zebra patterns or
similar were present. For example, frequency drifts with opposite directions
within a single stripe have been observed earlier \citep{Zlotnik15}, suggesting
a nonuniform coronal magnetic trap.

\section{Conclusions}

We have presented here observations of three radio type~IV burst events that
occurred on three separate days but originated from the same active region.
In the first event the source region was located near the solar disk center,
in the second event near W30, and in the third event near the west limb.
Our analysis shows that in the last two events the type IV bursts experienced
partial cut-offs in their emission. These two bursts were estimated to be
the moving type, i.e., associated with upward-moving CME structures instead
of stationary coronal loops. 

We conclude that partial cut-offs in IP type IV emission can appear also in
moving type IV bursts, which are not supposed to be directive like stationary
type IV bursts. The reason for the reduced emission could be absorption or
suppression of emission toward certain viewing angles, caused by increased
plasma density near CME flank and bow shock regions.
High densities can be found especially in streamer regions, and streamers
were observed near the CMEs in all three events.
However, as the type II burst frequencies are much lower than those of the
type IV bursts, they should not be able to block the radio propagation by
means of reflection. As the type IV burst emission mechanism - or mechanisms -
are not that well known either, this is clearly a topic that needs to
be studied further.

If radio type IV burst emission can be observed only if the burst source
is located near the solar disk center, this could be used to estimate
the CME propagation direction. However, this is not possible if the observed
directivity is caused by absorption/blocking of the type IV emission toward
certain directions. Instead, the effect will tell if high density,
shock-related regions are formed within CME structures, and where they
may be located.\\

\noindent{\bf Acknowledgements}

We thank all the individuals who have contributed in creating and updating
the various solar event catalogues and databases. We are especially grateful
to Stephen White for preparing and providing us the Green Bank Radio Burst
Spectrometer data. We thank the RSDB service at LESIA/USN (Observatoire de Paris)
for making the Nancay Radioheliograph data available. The CME catalog is
generated and maintained at the CDAW Data Center by NASA and the Catholic
University of America in cooperation with the Naval Research Laboratory.


\end{document}